\documentclass[runningheads]{llncs}

\usepackage[T1]{fontenc}
\usepackage{amsmath, array}
\usepackage{graphicx,verbatim}

\begin{document}
\title{Imagining Alternatives: Towards High-Resolution 3D Counterfactual Medical Image Generation via Language Guidance}

\titlerunning{Towards Language-Guided 3D Counterfactual Medical Image Generation}

\author{Mohamed Mohamed\thanks{Corresponding author: \email{mohamed.mohamed5@mail.mcgill.ca}}\inst{1,2} \and
Brennan Nichyporuk\inst{1,2} \and
Douglas L. Arnold\inst{1} \and \\
Tal Arbel\inst{1,2}}

\authorrunning{M. Mohamed et al.}

\institute{
McGill University \and
Mila – Quebec AI Institute}

\maketitle    

\begin{abstract}
Vision-language models have demonstrated impressive capabilities in generating 2D images under various conditions; however the impressive performance of these models in 2D is largely enabled by extensive, readily available pretrained foundation models. Critically, comparable pretrained foundation models do not exist for 3D, significantly limiting progress in this domain. As a result, the potential of vision-language models to produce high-resolution 3D counterfactual medical images conditioned solely on natural language descriptions remains completely unexplored. Addressing this gap would enable powerful clinical and research applications, such as personalized counterfactual explanations, simulation of disease progression scenarios, and enhanced medical training by visualizing hypothetical medical conditions in realistic detail. Our work takes a meaningful step toward addressing this challenge by introducing a framework capable of generating high-resolution 3D counterfactual medical images of synthesized patients guided by free-form language prompts. We adapt state-of-the-art 3D diffusion models with enhancements from Simple Diffusion and incorporate augmented conditioning to improve text alignment and image quality. To our knowledge, this represents the first demonstration of a language-guided native-3D diffusion model applied specifically to neurological imaging data, where faithful three-dimensional modeling is essential to represent the brain’s three-dimensional structure. Through results on two distinct neurological MRI datasets, our framework successfully simulates varying counterfactual lesion loads in Multiple Sclerosis (MS), and cognitive states in Alzheimer’s disease, generating high-quality images while preserving subject fidelity in synthetically generated medical images. Our results lay the groundwork for prompt-driven disease progression analysis within 3D medical imaging. Our code is accessible through the project website https://lesupermomo.github.io/imagining-alternatives/.
\end{abstract}

\section{Introduction}
Vision-language models have emerged as transformative tools for bridging textual and visual modalities, enabling powerful image synthesis and editing capabilities. Models such as DALL-E~\cite{ramesh2021zero} and Stable Diffusion~\cite{rombash2022latent} exemplify these advancements by producing and modifying high-quality images directly from natural language prompts. The impressive performance of these models in 2D is largely enabled by extensive, readily available pretrained foundation models. Recent innovations, including latent diffusion techniques~\cite{rombash2022latent} and instruction-driven image editing methods like InstructPix2Pix~\cite{brooks2023instructpix2pix}, have significantly enhanced the ability of these methods to generate photorealistic and semantically precise imagery. Although foundational 2D diffusion models have influenced developments in 3D-aware natural image applications, such as object editing~\cite{michel2023object}, 3D human animation~\cite{wang2022humanise}, and detailed face generation driven by language attributes~\cite{wu2023high}, existing vision-language methods for 3D predominantly focus on external object surfaces rather than the detailed internal volumetric structures crucial in medical imaging. Critically, comparable pretrained foundation models do not exist for 3D, significantly limiting progress in this domain.

\begin{figure}[t]
    \centering
    \includegraphics[width=0.85\linewidth]{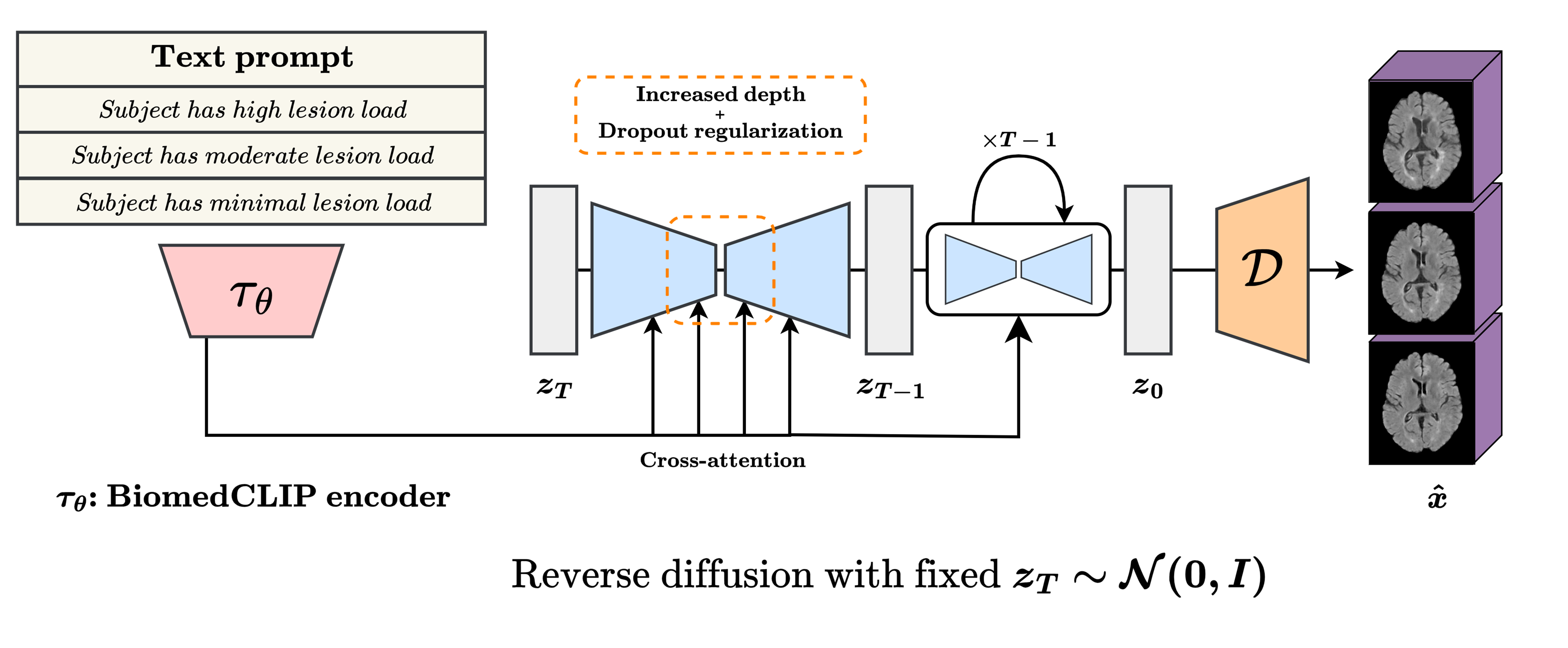}
    \caption{Proposed Framework. A pretrained BiomedCLIP text encoder encodes the text prompt (e.g.\ ``Subject has high lesion load'') as conditioning for the diffusion model. During inference, the model generates counterfactuals by sampling from the same fixed noise while varying the text condition.}
    \label{fig:method}
\end{figure}

Medical imaging, particularly neurological imaging, requires accurate volumetric modeling to faithfully represent anatomical complexities inherent in the brain’s three-dimensional structure. Recent advances, such as Text2CT~\cite{guo2025text2ct} and MedSyn~\cite{xu2024medsyn}, have successfully generated anatomically plausible 3D volumes guided by structured medical texts. However, these methods primarily produce novel synthetic scans rather than addressing the clinically important ability to create controlled, counterfactual modifications to existing patient-specific images. Moreover, current methods remain predominantly unconditional or constrained by limited numerical or categorical variables, such as patient age or diagnostic labels~\cite{friedrich2024wdm}. Furthermore, the application of 3D generative models in medical imaging faces substantial challenges, including higher data demands, increased computational complexity, and limited availability of curated volumetric datasets compared to their 2D counterparts.

Addressing this gap, we introduce the first framework capable of generating high-resolution, text-guided 3D counterfactual medical images of \emph{synthetic subjects}, allowing researchers to explore nuanced \textit{what-if} disease-progression scenarios. Unlike prior approaches reliant on slice-based sequential modeling, our approach employs a native-3D, language-guided diffusion backbone, representing the first demonstration of such a model specifically tailored to neurological imaging. We build upon state-of-the-art 3D diffusion architectures, specifically leveraging enhancements from Simple Diffusion~\cite{hoogeboom2023simple}, such as targeted bottleneck expansion and dropout regularization, alongside incorporating additional cross-attention modules at higher-resolution scales as well as medically-informed semantic embeddings derived from BiomedCLIP~\cite{zhang2024biomedclip}. We demonstrate that language-guided wavelet-based diffusion models (WDM), which operate directly in voxel space and support fine-grained counterfactual edits, offer superior subject preservation and achieve text alignment on par with or exceeding that of latent diffusion models. This makes WDM a compelling choice for 3D counterfactual image editing tasks requiring subtle anatomical modifications. In parallel, we enhance the MAISI~\cite{10943915} 3D latent diffusion framework by incorporating a Rectified Flow noise schedule (MAISI RFlow), leading to significant gains in image fidelity and anatomical consistency compared to traditional linear scheduling (MAISI Linear). Notably, MAISI RFlow achieves subject-preservation and text-alignment scores approaching those of WDM, while reducing memory and compute requirements by 65\%.

Through extensive experiments on brain MRI datasets representing patients with neurological diseases, our method successfully simulates diverse counterfactual lesion distributions in \emph{synthetic} Multiple Sclerosis (MS) patients and varied cognitive states in \emph{synthetic} Alzheimer’s disease patients. Separately, we show how textual alignment and anatomical fidelity can be balanced using classifier-free guidance~\cite{ho2022classifierfree}, illustrating the explicit trade-offs inherent to conditional generative modeling. This framework thus lays essential groundwork for leveraging vision-language models in generating counterfactual images of \emph{real} patients, paving the way for personalized disease modeling, realistic clinical image synthesis, and improved medical education in the domain of 3D medical imaging.

\section{Methodology}
\begin{figure}[t]
  \centering
  \includegraphics[width=0.80\textwidth]{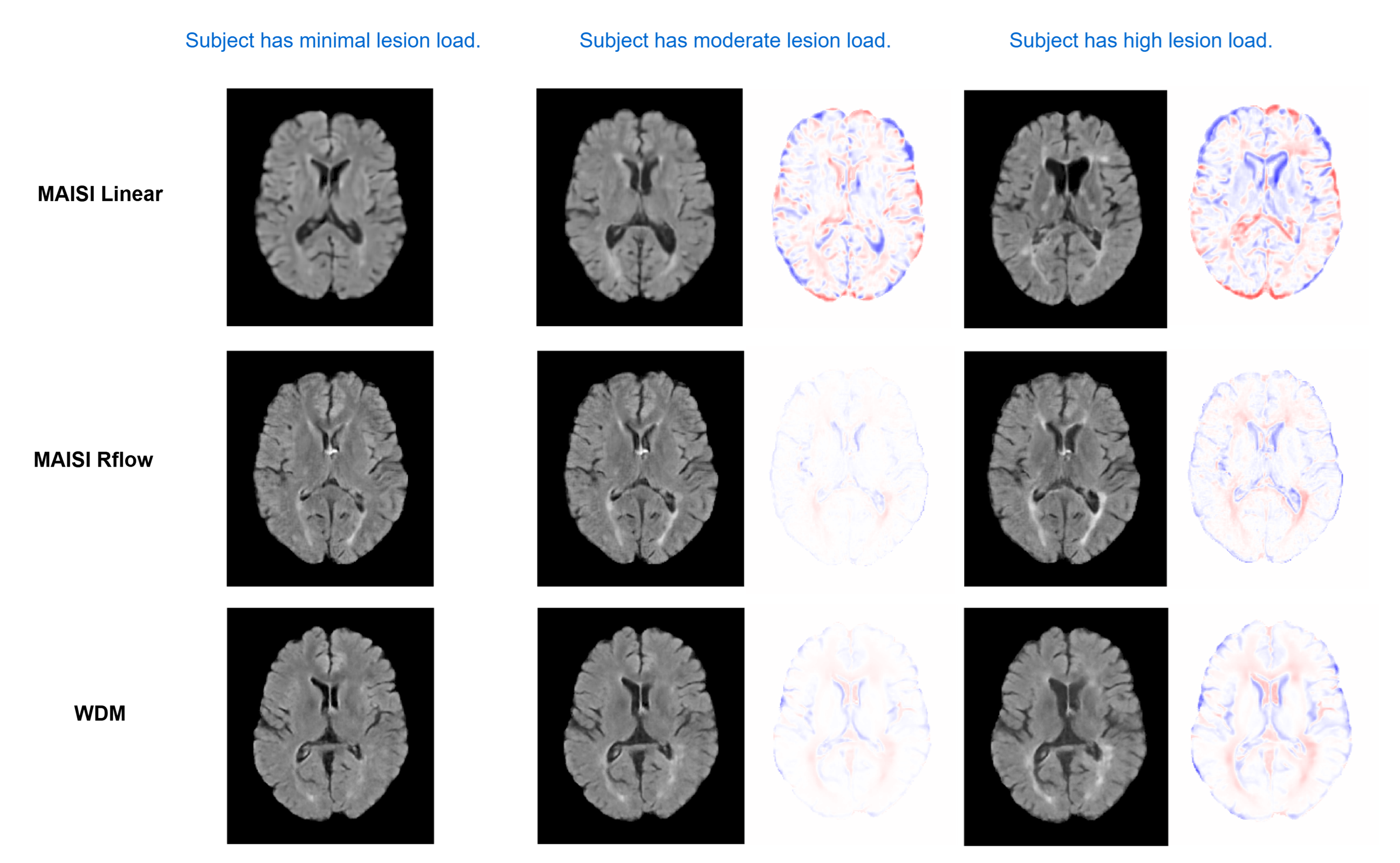}
  \caption{Qualitative comparison of generated counterfactuals for synthesized subjects on the MS dataset for different lesion loads.}
  \label{fig:qualitative_loris}
\end{figure}

In this section we first describe the process of creating text prompts from image-derived features or clinical variables. Next, we describe our language-guided 3D diffusion model architecture and training procedure. Finally, we describe the process of generating language-guided 3D counterfactuals of synthetic patients.

\subsection{Creating Text Prompts}
To maximize the effectiveness of text conditioning during inference, we carefully select textual prompts that correspond to prominent and easily distinguishable features in the MRI scans—such as lesion burden or global atrophy patterns. These attributes manifest clearly in the imaging data and help ensure that the model can meaningfully condition on the provided text.

\subsection{Language-Guided 3D Diffusion Model}
We adopt three architectural enhancements inspired by Simple Diffusion~\cite{hoogeboom2023simple}. (i) The U-Net bottleneck is deepened with additional residual blocks — an effective way to add capacity with minimal memory/compute overhead. (ii) Targeted dropout is applied in the lower-resolution layers (including the bottleneck) to regularize training and boost image fidelity. (iii) We maximize the use of cross-attention layers at multiple scales, enabling text embeddings to steer both coarse and fine features. During training, we replace the prompt with the null text 20\% of the time, enabling classifier-free guidance at inference.

For the noise schedule, we compare the linear schedule of (\textbf{WDM})~\cite{friedrich2024wdm} and the MAISI baseline (\textbf{MAISI Linear}) with a rectified-flow variant (\textbf{MAISI Rflow}) that follows the straightened diffusion trajectories proposed by Liu et al.~\cite{liu2023rectified}. Rectified flow theoretically yields higher sample quality and faster inference by learning an optimal-transport ODE between data and noise.

\begin{figure}[t]
  \centering
  \includegraphics[width=0.80\textwidth]{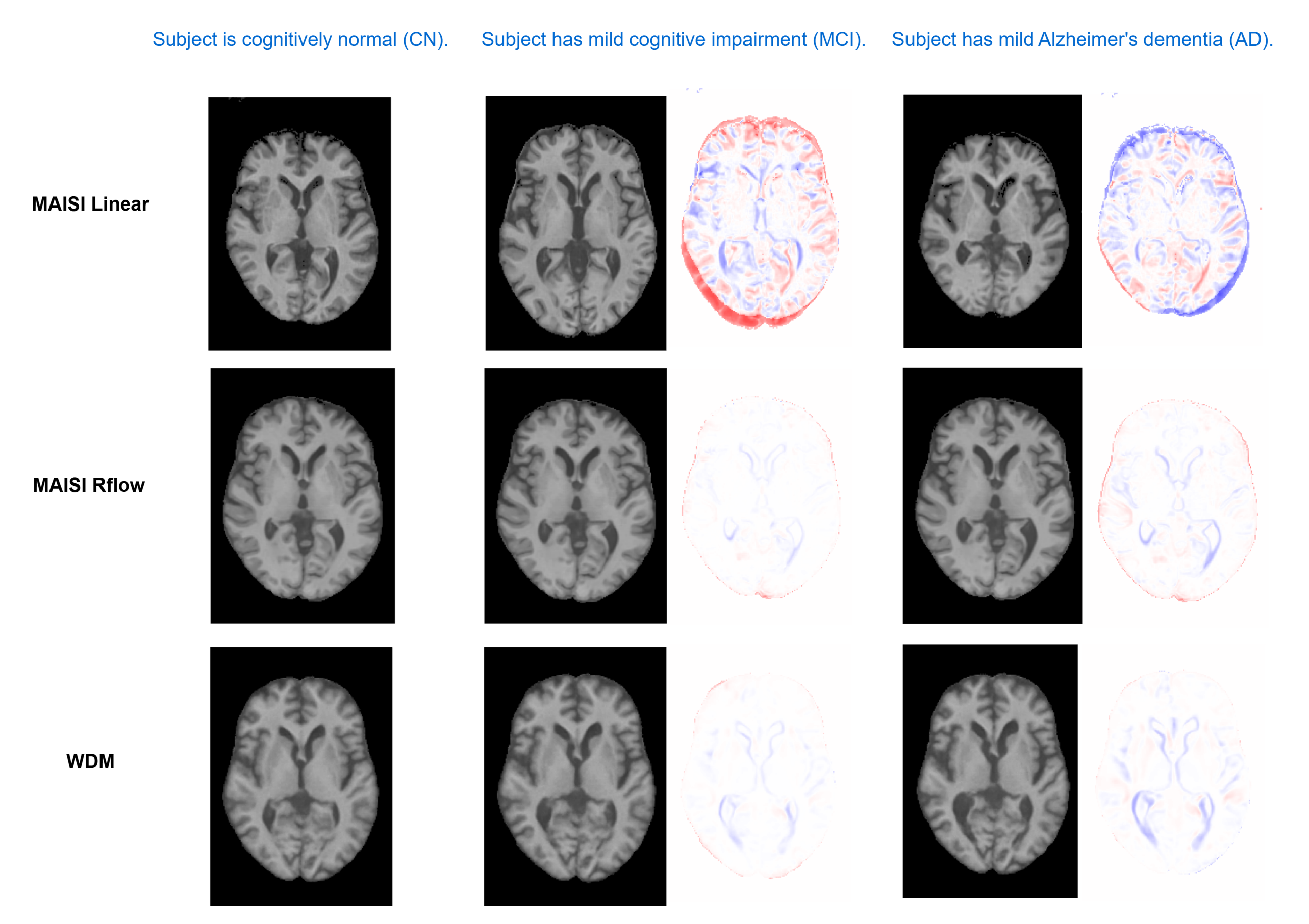}
  \caption{Qualitative comparison of generated counterfactuals for synthesized subjects on the ADNI dataset for different cognitive states.}
  \label{fig:qualitative_adni}
\end{figure}

\subsection{Generating Language-Guided 3D Counterfactuals}
Our counterfactual generation method leverages a text-conditioned diffusion model at inference time to produce samples that differ only in the condition of interest. During sampling, we start from a fixed source of Gaussian noise and generate images using different text prompts. Formally, let $x_T \sim \mathcal{N}(0,I)$ be a random latent (at the highest diffusion timestep $T$). We generate one image $x^{(a)}_0 = D_{\theta}(x_T, y_a)$ conditioned on text prompt $y_a$ (e.g. “Subject has low lesion load.”), and another image $x^{(b)}_0 = D_{\theta}(x_T, y_b)$ from the same noise $x_T$ but with an alternative prompt $y_b$ (e.g. “Subject has high lesion load.”). Here $D_{\theta}(\cdot, y)$ denotes the diffusion model’s sampling function decoding noise into an image given condition $y$. This yields counterfactual images that share the same underlying subject identity while reflecting different clinically meaningful states.

\section{Experiments and Results}

\subsection{Datasets \& Implementation details}

We evaluate our approach on two 3D MRI datasets: (1) a proprietary, multi-center dataset of 7107 FLAIR scans from 10 multiple sclerosis (MS) clinical trials, where subjects are grouped by lesion volume into minimal (0–10 mL), moderate (10–25 mL), and high (>25 mL) categories, which are used as text conditions; and (2) the publicly available ADNI dataset comprising 1874 T1-weighted scans labeled as cognitively normal (CN), mild cognitive impairment (MCI), or Alzheimer's disease (AD). Both datasets are split into 70/15/15 for training, validation, and testing. To mitigate class imbalance, we apply weighted sampling to emphasize underrepresented conditions (e.g., high lesion load or AD).

Models are implemented in PyTorch and trained for 1 million steps with a batch size of 1. Latent diffusion models use a DDPM sampler with 1000 denoising steps, while the WDM variant achieved best performance (FID, subject preservation, text alignment) using a DDIM sampler with 25 steps.

\subsection{Experiments and Metrics}
For each model–dataset pair we synthesize 1,000 volumes—333 per prompt—then generate \emph{medium‑} and \emph{high‑level} counterfactuals from every \emph{low‑level} baseline (e.g., low vs. moderate/high lesion burden, cognitively normal vs. MCI/AD). Qualitative examples with difference maps appear in Figure~\ref{fig:qualitative_loris} and Figure~\ref{fig:qualitative_adni}.

\begin{table}[t]
  \centering
  \caption{Performance of the MAISI models and the WDM model for both the MS and ADNI dataset, across image quality, subject preservation, and text alignment accuracy.}
  \label{table:combined_metrics}
  \setlength{\tabcolsep}{2pt}
  \begin{tabular}{c|cc|cc|c}
    \hline
    & \multicolumn{2}{c|}{\textbf{Quality and Diversity}} 
    & \multicolumn{2}{c|}{\textbf{Subject Preservation}} 
    & \multicolumn{1}{c}{\textbf{Text Alignment}} \\
    \cline{2-6}
    \textbf{Model} & FID $\downarrow$ & MS-SSIM $\downarrow$ 
    & MS-SSIM $\uparrow$ & PSNR (dB) $\uparrow$ 
    & Accuracy (\%) $\uparrow$ \\
    \hline
    \multicolumn{6}{c}{\textbf{MS Dataset}} \\
    \hline
    MAISI Linear  & 0.1734 & \textbf{0.8626} & 0.8614 & 21.18 & 92.64 \\
    MAISI Rflow   & 0.1625 & 0.8715 & 0.9638 & 27.92 & 89.19 \\
    WDM           & \textbf{0.1622} & 0.8684 & \textbf{0.9680} & \textbf{28.47} & \textbf{94.58} \\
    \hline
    \multicolumn{6}{c}{\textbf{ADNI}} \\
    \hline
    MAISI Linear  & 0.1752 & 0.8362 & 0.8635 & 22.16 & 62.26 \\
    MAISI Rflow   & 0.1690 & 0.8357 & 0.9864 &\textbf{ 34.86} & \textbf{72.01}\\
    WDM           & \textbf{0.1647} & \textbf{0.7485} & \textbf{0.9895} & 34.77 & 71.17\\
    \hline
  \end{tabular}
\end{table}

We evaluate three complementary goals.
\textbf{(i) Image quality:} realism and diversity are captured by FID (Med3D features~\cite{chen2019med3d}; lower is better) and mean MS‑SSIM across 1,000 generated samples (lower = more varied).
\textbf{(ii) Subject preservation:} MS‑SSIM and PSNR between each factual image and its counterfactual quantify how faithfully anatomy is retained (higher is better).
\textbf{(iii) Text alignment:} a 3‑D DenseNet‑121 classifier is trained to judge whether the generated volume reflects the requested clinical state; its accuracy on held‑out \emph{real} scans—96.5\% for MS (low vs. high lesion load) and 83.3\% for ADNI (CN vs. AD)—provides a strong baseline, so any accuracy drop on counterfactuals reliably indicates misalignment.

\subsection{Results}
In this section, we present qualitative and quantitative results for WDM, MAISI Rflow, and MAISI Linear. Finally, we demonstrate how classifier-free guidance can be used to trade off text-alignment and subject-preservation in the context of counterfactual generation of synthetic subjects.

\textbf{Qualitative Counterfactual Generation.}
Figures~\ref{fig:qualitative_loris} and~\ref{fig:qualitative_adni} show counterfactual MRIs generated by our latent diffusion and adapted WDM models. On the MS dataset, MAISI Linear yields low‑quality images with off‑target changes, whereas MAISI Rflow and WDM selectively increase lesion load while preserving anatomy. The same holds for ADNI: only Rflow and WDM convincingly depict Alzheimer‑related cortical atrophy and ventricular enlargement. These results confirm that the rectified‑flow noise schedule boosts image fidelity and counterfactual control in latent diffusion models, and that our approach generalizes to voxel‑space diffusion via WDM.

\textbf{Quantitative Evaluation.}
Quantitative results summarizing image quality, subject preservation, and text alignment metrics are reported in Table~\ref{table:combined_metrics}. Language-guided wavelet-based diffusion models (WDM), which operate directly in voxel space and support fine-grained counterfactual edits, offer superior subject preservation on par with or exceeding that of latent diffusion models. They also achieve comparable or superior text alignment, making them an ideal model for subject preserving counterfactual edits. Furthermore, they achieve comparable or superior FID scores on both datasets, indicating superior realism and image quality. In terms of diversity, the WDM baseline yields either the lowest or comparable MS-SSIM scores across generated samples, highlighting superior sample diversity. The MAISI RFlow model achieves subject preservation performance that approaches, or in some cases exceeds that, of the WDM model, but requires 65\% less training time, making it a lightweight counterfactual model that can be trained and experimented with quickly in research contexts. Although the MAISI linear model achieved the best sample diversity for the MS dataset, this appeared to stem from, at least in part, exaggerated and unrealistic changes across text prompts, as demonstrated by the worst FID score on the MS dataset. On all other metrics, MAISI linear model almost always achieved the worst performance, making it a poor candidate for fine-grained counterfactual generation in the context of neurological diseases.

\textbf{Classifier-Free Guidance Ablation.}
Table~\ref{table:guidance_ablation} presents the effects of varying the classifier-free guidance scale ($w$) using MAISI Rflow on the MS dataset. Overall, higher guidance scales significantly enhance prompt fidelity but tend to reduce anatomical consistency. This demonstrates how classifier-free guidance can be tuned to balance text alignment and subject preservation, depending on the researcher's priorities.

\begin{table}[t]
  \centering
  \caption{Ablation of guidance scale (MAISI Rflow on MS).}
  \label{table:guidance_ablation}
  \setlength{\tabcolsep}{6pt}
  \begin{tabular}{c|cccc}
    \hline
    \textbf{Guidance} & \textbf{Text Alignment (\%) $\uparrow$} & \textbf{Subject MS-SSIM $\uparrow$} & \textbf{PSNR (dB) $\uparrow$}  \\
    \hline
    No CFG & 89.17 & 0.964 & 27.92 \\
    \hline
    0.5 & 70.08 & 0.985 & 32.30  \\
    1 & 85.96 & 0.964 & 27.89  \\
    2 & 99.39 & 0.926 & 24.39  \\
    3 & 100 & 0.898 & 22.75 \\
    \hline
  \end{tabular}
\end{table} 

\section{Conclusion}
In this work, we introduced a novel vision-language framework designed specifically for generating high-resolution, text-guided 3D counterfactual medical images of synthetic neurological subjects. Our approach addresses critical limitations of existing methods by integrating advanced diffusion architectures with medically-informed semantic embeddings derived from BiomedCLIP. The results demonstrate that our language-guided wavelet-based diffusion model (WDM), operating directly in voxel space, delivers superior subject preservation, image quality, and text alignment compared to conventional latent diffusion approaches. Additionally, the MAISI RFlow model, which incorporates a Rectified Flow noise schedule, significantly improves anatomical consistency and image fidelity while achieving computational efficiency. Qualitative and quantitative analyses clearly indicate the effectiveness of these models in simulating nuanced disease-progression scenarios in synthetic Multiple Sclerosis and Alzheimer's patients, and our ablation studies on classifier-free guidance underscore the explicit trade-offs between prompt fidelity and anatomical accuracy. Ultimately, our contributions lay essential groundwork for future extensions towards generating personalized counterfactual medical images from real patient data, with potential implications for enhanced clinical decision-making, personalized disease modeling, and medical education within 3D neurological imaging.

\begin{credits}
\subsubsection{\ackname} This investigation was supported by the International Progressive Multiple Sclerosis Alliance (PA-1412-02420) and the companies who generously provided the data: Biogen, BioMS, MedDay, Novartis, Roche/Genentech, and Teva; the MS Society of Canada; the Natural Sciences and Engineering Research Council of Canada; the Canadian Institute for Advanced Research (CIFAR) Artificial Intelligence Chairs program; Calcul Quebec; the Digital Research Alliance of Canada; and Mila - Quebec AI Institute. Data collection and sharing for the Alzheimer's Disease Neuroimaging Initiative (ADNI) dataset is funded by the National Institute on Aging (National Institutes of Health Grant U19AG024904).

\subsubsection{\discintname} The authors have no competing interests to declare.
\end{credits}

\bibliographystyle{splncs04}
\bibliography{paper-0027} 

\end{document}